\documentclass[floats,floatfix,showpacs,amssymb,prd,twocolumn,superscriptaddress,nofootinbib,nolongbibliography,reprint]{revtex4-2}

\usepackage{amsmath}
\usepackage{amsfonts}
\usepackage{amssymb}
\usepackage{booktabs}
\usepackage[T1]{fontenc}
\usepackage{graphicx}
\usepackage[dvipsnames, usenames]{xcolor}
\definecolor{linkcolor}{rgb}{0.0,0.3,0.5}
\usepackage[unicode, colorlinks=true, linkcolor=linkcolor, citecolor=linkcolor, filecolor=linkcolor,urlcolor=linkcolor, pdfusetitle]{hyperref}
\usepackage{orcidlink}
\usepackage{siunitx}

\DeclareSIUnit\sun{M_{\odot}}

\newcommand{\milan}{Dipartimento di Fisica ``G. Occhialini'', 
Universit\`a degli Studi di Milano-Bicocca, Piazza della Scienza 3, 20126 Milano, Italy}
\newcommand{\infn}{INFN, Sezione di Milano-Bicocca, 
Piazza della Scienza 3, 20126 Milano, Italy}

\newcommand{\lalsuite}{\textsc{lalsuite}}

\newcommand{\amp}{A}
\newcommand{\phase}{\varphi}

\newcommand{\cohsum}{\mathcal{C}}
\newcommand{\Dir}{\mathfrak{Dir}}
\newcommand{\Fre}{\mathfrak{Fre}}
\newcommand{\scri}{A}

\newcommand{\jax}{\textsc{jax}}
\newcommand{\phenomt}{\textsc{IMRphenomT}}
\newcommand{\sfts}{\textsc{sfts}}

\newcommand{\chirpM}{\mathcal{M}}

\newcommand{\Fc}{F_{\times}}
\newcommand{\Fp}{F_{+}}
\newcommand{\Nsft}{N_{\mathrm{SFT}}}

\newcommand{\scal}[2]{\left\langle #1 , #2 \right\rangle}
\newcommand{\Sn}{S_{\mathrm{n}}}
\newcommand{\tdscal}[2]{\left(\left[#1, #2\right]\right)} 

\newcommand{\Tsft}{T_{\mathrm{SFT}}}
\newcommand{\proj}{\Lambda}

\begin{document}

\title{
    Scalable data-analysis framework for long-duration gravitational waves\\from
    compact binaries using short Fourier transforms
}

\author{Rodrigo Tenorio$\,$\orcidlink{0000-0002-3582-2587}}
\email{rodrigo.tenorio@unimib.it}
\affiliation{\milan}
\affiliation{\infn}
\author{Davide Gerosa$\,$\orcidlink{0000-0002-0933-3579}}
\affiliation{\milan}
\affiliation{\infn}

\begin{abstract}
    We introduce a framework based on short Fourier transforms (SFTs) to analyze long-duration gravitational 
    wave signals from compact binaries. Targeted systems include binary neutron stars observed by third-generation
    ground-based detectors and massive black-hole binaries observed by the LISA space mission. 
    In short, ours is an extremely fast, scalable, and parallelizable implementation of the gravitational-wave inner product,
    a core operation of gravitational-wave matched filtering.
    By operating on disjoint data segments, SFTs allow for efficient handling of noise
    non-stationarities, data gaps, and detector-induced signal modulations.
    We present a pilot application to early warning problems in both ground- and space-based next-generation
    detectors. 
    Overall, SFTs reduce the computing cost of evaluating an inner product by three to five orders
    of magnitude, depending on the specific application, with respect to a non-optimized approach.
    We release public tools to operate using the SFT framework, including a vectorized and hardware-accelerated
    re-implementation of a time-domain waveform.
    The inner product is the key building block of all gravitational-wave data treatments; by speeding up this 
    low-level element so massively, SFTs provide an extremely promising solution for current and future gravitational-wave
    data-analysis problems.
\end{abstract}

\maketitle

\section{Introduction}

The detection of gravitational-wave signals (GWs) from compact object coalescences (CBCs),
such as binary black holes (BBHs) or neutron stars (BNS) is now routine~\cite{KAGRA:2021vkt, Nitz:2021zwj,Wadekar:2023gea, Koloniari:2024kww}.
As observed by the current network of ground-based detectors (LIGO~\cite{LIGOScientific:2014pky}, 
Virgo \cite{VIRGO:2014yos}, and KAGRA~\cite{KAGRA:2018plz}),
these signals sweep the audible band of the GW spectrum with durations ranging from a fraction
of a second for BBHs to up to a few minutes for BNSs.

The duration of a GW signal is a function of both the intrinsic properties of the system 
---notably  its chirp mass--- and the minimum sensitive frequency of the detector~\cite{Maggiore:2007ulw}.
While LIGO/Virgo/KAGRA are sensitive to frequencies $\gtrsim \SI{10}{\hertz}$,
future ground-based detectors, such as Einstein Telescope (ET)~\cite{ET:2019dnz} and
Cosmic Explorer (CE)~\cite{Reitze:2019iox} are expected to bring this limit down to
\mbox{$\gtrsim\SI{1}{\hertz}$}. With  this extended frequency range, BNSs will stay in band
for up to a week. The LISA space mission~\cite{LISA:2024hlh} will  soon detect GW sources at mHz frequencies. These include massive BBHs with masses on
the order of $10^{6} \mathrm{M}_\odot$~\cite{Katz:2019qlu}, which will be observable for weeks;
other sources such as extreme mass-ratio inspirals~\cite{Babak:2017tow} and galactic white dwarf
binaries~\cite{Littenberg:2018xxx} will be observable for years.

Such long-duration signals challenge the current approach to CBC data
analysis~\cite{Allen:2005fk, Harry:2010fr, Veitch:2014wba, Usman:2015kfa, Ashton:2018jfp, Chu:2020pjv,
CANNON2021100680, Huang:2024mwa} and require prompt attention to fully exploit the scientific potential
of future GW observatories.

The analysis of CBC signals is grounded on matched filtering, which extensively uses
the inner product~\cite{1987thyg.book..330T,Finn:1992wt}
\begin{equation}
    \scal{x}{h}
    = 4 \, \mathrm{Re} \int_{0}^{\infty} \mathrm{d}f \frac{\tilde{x}^{*}(f) \tilde{h}(f)}{\Sn(f)} \,,
    \label{eq:fd_product}
\end{equation}
where $x$ is the observed data, $h$ is a signal (both here expressed in the frequency domain),
and $\Sn$ is the single-sided power spectral density of the noise. Equation~\eqref{eq:fd_product}
is the main entry point of data and waveform templates to an analysis. Any significant improvement
on its implementation will significantly influence the future of the field in terms of data-processing strategies,
waveform-model implementations, and overall pipeline architecture.

The frequency domain formulation of Eq.~\eqref{eq:fd_product} is convenient for short-duration signals,
as $\tilde{h}(f)$ can be efficiently evaluated with waveform approximants
(e.g.~Refs.~\cite{Varma:2019csw,Pratten:2020fqn,Ramos-Buades:2023ehm,Nagar:2024dzj}), 
and different frequency values are taken as independent, thus simplifying the calculation.
Applying Eq.~\eqref{eq:fd_product} to signals longer than a few minutes, on the other hand, poses significant challenges:
\begin{enumerate}
\item The evaluation of $\tilde{h}(f)$ is complicated due to the amplitude
    and frequency modulations imprinted by the detector.  While these can be 
    approximated in Fourier domain using closed-form 
    expressions~\cite{Marsat:2020rtl,Chen:2024kdc}, their cost dominates over 
        that of generating the waveform.
\item GW detectors are not always
    operational~\cite{LIGO:2021ppb,LIGO:2024kkz}, causing gaps in the observed data stream; this further 
    complicates the correct and efficient computation of frequency-domain quantities. 
\item Long-duration non-stationarities of the detector noise 
    are difficult to model in the frequency domain~\cite{Kumar:2022tto}.
\end{enumerate}

Recently proposed accelerated likelihoods for long waveforms such as
multi-banding~\cite{Vinciguerra:2017ngf,Garcia-Quiros:2020qlt,Morisaki:2021ngj}
relative binning~\cite{Zackay:2018qdy,Leslie:2021ssu,Krishna:2023bug}, 
or heterodyning~\cite{Cornish:2021lje} do not address these three key challenges 
in GW data analysis~\cite{Baker:2025taj}. 
Other solutions include time-frequency formulations based on wavelets~\cite{Cornish:2020odn}.
While potentially effective, arbitrary functional bases tend to introduce
computational overheads whenever converting from purely time or frequency 
domains where both signal and detector properties tend to be more naturally defined.

The analysis of long-duration signals with a narrow spectral structure has been thoroughly studied
in the context of continuous gravitational-wave (CWs)
searches~\cite{Tenorio:2021wmz,Riles:2022wwz,Wette:2023dom}.
Leveraging CW search strategies~\cite{Williams:1999nt,Prix:2011qv,prixCFSv2,fasttracks},
we present a new, fast, scalable implementation of the GW inner product from 
Eq.~\eqref{eq:fd_product} based on short Fourier transforms (SFTs).
We develop the SFT formalism for the analysis of GWs from compact binaries
and present a pilot application applied to the early warning of both BNS signals in
third-generation ground-based detectors and massive BBH signals in LISA. 
The computational advantages of SFTs provide a promising approach
for low-latency GW analyses targeting multi-messenger science,
where every instant gained by GW data-analysis corresponds to new and
potentially ground-breaking science on the electromagnetic
side~\cite{LIGOScientific:2017vwq, Mangiagli:2022niy}.

Our new approach addresses the three key challenges identified above by construction.

\begin{figure}
    \includegraphics[trim=1.5cm 0 3cm 1.9cm,clip=true,width=\columnwidth]{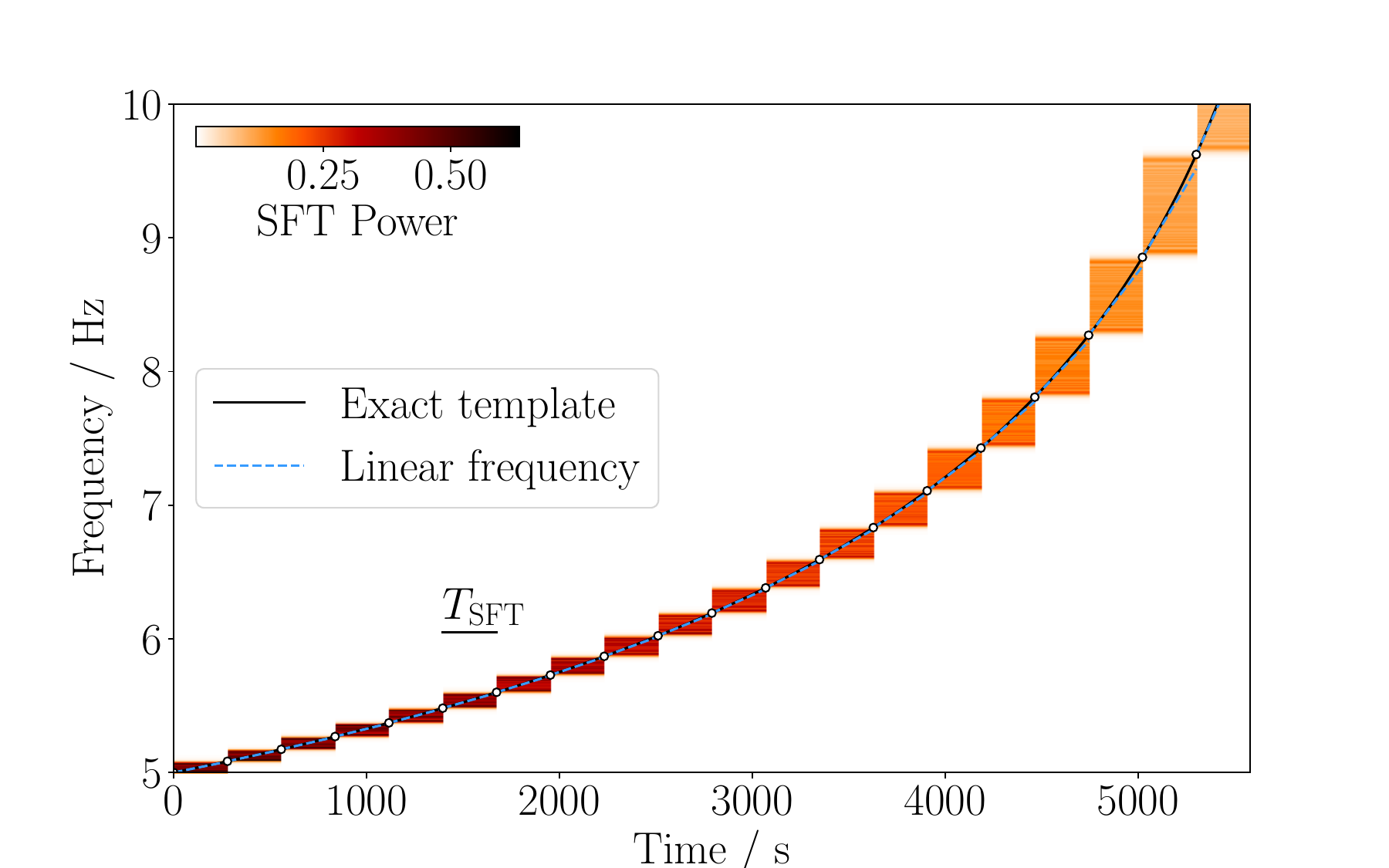}
    \caption{
        Spectrogram of a GW signal compatible with the inspiral of a
        \mbox{$(1.4 - 1.4) \, \mathrm{M}_{\odot}$} BNS. 
        This corresponds to the squared absolute value of a set of SFTs. 
        Colored regions correspond to frequency ranges where the GW signal is present.
        As the signal evolves, the frequency increases at a faster rate, broadening the colored frequency ranges.
        The signal frequency evolution, shown as a solid black curve, 
        is computed using a state-of-the-art time-domain waveform model. 
        The approximated linear frequency, shown as a dashed blue curve, is computed at the beginning of each SFT (white circles).
        The inner product can be efficiently computed by folding in the complex amplitude of
        signal-dominated frequency regions in each SFT and accumulating the result over time.
        We used $\Tsft = \SI{280}{\second}$, which is unrealistically large (see Sec.~\ref{sec:pilot}),
        to better illustrate the basic idea. On a realistic application the highlighted regions become
        an order of magnitude narrower.
    }
    \label{fig:example}
\end{figure}

The use of SFTs is schematically shown in Fig.~\ref{fig:example}. 
First, the original dataset is split into ``short'' time segments which then get Fourier transformed.
Each of this segments corresponds to an SFT. The set of all SFTs forms a complex-valued 2D array
with a time axis and a frequency axis, whose absolute value squared is to be identified with a spectrogram.
At each SFT, the GW signal under study is spread across a narrow frequency range
(colored fringes in Fig.~\ref{fig:example}). In particular, Eq.~\eqref{eq:fd_product}
can be computed by accumulating SFT values at the frequency bins around the signal'
s frequency weighed by an appropriate kernel. 
Since SFTs are treated independently, the analysis can easily
deal with noise non-stationarities, time-dependent modulations induced by the detector,
and missing data periods.

We further exploit the slow frequency evolution of long-duration GWs
to massively downsample the dataset using SFTs once and for all for a given analysis.
This reduces the memory footprint of the analysis significantly reducing the computing 
cost and allowing for batch-evaluating likelihoods using hardware accelerators such 
as GPUs. 

The required tools to operate using the SFTs framework are released as part of an
accompanying open-source software package, \sfts{}~\cite{sfts}. We note that
the Fourier-transform convention chosen in this work are consistent with
the data contained in the \texttt{.sft} files used by the LIGO/Virgo/KAGRA Collaboration~\cite{SFT_format}.

Owing to its simplicity, scalability, and computational efficiency, our SFT approach paves the way for a
new paradigm in GW data analysis, offering a powerful framework to tackle challenges to come. 

\section{Gravitational-wave data analysis}

\subsection{Strain and detector projection}

Let us first review the basic tools for the analysis of GW data.
GW data analysis operates with a log-likelihood~\cite{Jaranowski:1998qm}
\begin{equation}
    \mathcal{L}(\theta) = \scal{x}{h(\theta)} - \frac{1}{2}\scal{h(\theta)}{h(\theta)} \,,
    \label{eq:logl}
\end{equation}
where $x$ is the observed data and $h(\theta)$ is a signal template characterised
by a set of parameters $\theta$. This assumes that (i) the noise is Gaussian and (ii)
the power spectral density of the noise can be reasonably estimated. 
The specific formulation of $\scal{\cdot}{\cdot}$ in Eq.~\eqref{eq:logl} depends on the
application, namely parameter-estimation routines or detection statistics for search
pipelines~\cite{Allen:2005fk, Jaranowski:1998qm, Prix:2011qv,Capano:2016dsf,
thrane2019introduction, Christensen:2022bxb,Chua:2022ssg,Covas:2022xyd,Prix:2024xpl}.

In this paper,  we treat the problem of long-duration GW signals produced
by compact binaries. We consider
signals that start to be observed at an early stage of their inspiral, so that most of
their power is narrowly concentrated in the frequency domain at any given time.
Signals beyond this regime can still make use of the methods here discussed by excising
the inspiral from the rest of the waveform.

We describe a GW signal in terms of phase $\phase(t)$ and amplitude $A(t)$, which
in turn give rise to two polarizations $h_{+}(t)$ and $h_{\times}(t)$. 
As the signal reaches the detector, the observed strain can be expressed as~\cite{Jaranowski:1998qm}
\begin{equation}
    h(t) = \Fp(t) h_{+}(t) + \Fc(t) h_{\times}(t)
    \label{eq:strain}
\end{equation}
where 
\begin{equation}
    \begin{bmatrix}
        \Fp(t) \\ \Fc(t)
    \end{bmatrix} 
    = \sin\zeta \begin{bmatrix}
        \cos{2\psi} & \sin{2\psi} \\
        -\sin{2\psi} & \cos{2\psi}
    \end{bmatrix}
    \begin{bmatrix}
        a(t) \\ b(t)
    \end{bmatrix}
\end{equation}
are the polarization-dependent antenna patterns,
$\psi$ is the polarization angle, $a(t), b(t)$ are the
detector-dependent antenna pattern functions,
and $\zeta$ is the opening angle of the detector (e.g. $\zeta = 90^\circ$ for LIGO/Virgo/Kagra).
This description of $h(t)$ is appropriate under the so-called
long-wavelength approximation,
which is valid for current and future ground-based detectors below a few thousand kHz and for 
LISA up to a few mHz~\cite{Rakhmanov:2008is,Marsat:2020rtl,Virtuoso:2024kyp}.
Let us also define a projector
\begin{align}
    \proj_{t} &= \Fp(t) \scri_{+} \mathrm{Re} + \Fc(t) \scri_{\times} \mathrm{Im}
\end{align}
where $\scri_{+,\times}$ are a function of amplitude and orientation coefficients
so that 
\begin{equation}
    h(t) = \proj_{t} A(t) e^{i \phase(t)} \,.
\end{equation}
This projector allows us to abstract away any dependencies on the number of detectors, 
the number of modes, and the number of polarizations in the GW signal.

Throughout this work, we will take $\phase(t)$ and $A(t)$ be consistent with 
the $(2, \pm 2)$ GW mode of~\mbox{\phenomt{}}~\cite{Estelles:2020osj,Estelles:2020twz,
Estelles:2021gvs}. Our specific implementation~\cite{sfts}
is differentiable and GPU-vectorized using \jax{}~\cite{jax} as described in
Appendix~\ref{app:phenomt}.

\subsection{Detection statistics}

In the applications here presented, we are primarily concerned with GW searches.
These evaluate a detection statistic, which is usually derived from Eq.~\eqref{eq:logl},
on a set of waveform templates in order to determine the presence of a signal.

Current CBC analyses derive a detection statistic by maximizing Eq.~\eqref{eq:logl}
with respect to the initial GW phase and distance~\cite{Allen:2005fk,Usman:2015kfa,Capano:2016dsf}: 
\begin{align}
    \max_{\mathrm{distance}, \phase_{0}} \mathcal{L} &= 
    \frac{\scal{x}{h_{\mathrm{c}}}^2 + \scal{x}{h_{\mathrm{s}}}^2}{\scal{h_{\mathrm{c}}}{h_{\mathrm{c}}}}\,,
    \label{eq:det_short} \\
    h_{\mathrm{c}} &=  A(t) \cos\phase(t) \,, \\
    h_{\mathrm{s}} &= - A(t) \sin\phase(t) \,.
\end{align}
This approach is based on the fact that, for a short signal, the detector response
is constant and the two polarizations cannot be resolved. It also assumes that only the dominant GW mode is present.

For long-duration signals, the amplitude modulations imprinted by the detector
cannot be neglected and the two polarizations become distinguishable~\cite{Covas:2022xyd}.
This motivates the following parametrization for the observed strain:
\begin{equation}
    h(t) = \sum_{\nu = 0}^{3} c_{\nu} h_{\nu}(t) \,,
    \label{eq:det_long}
\end{equation}
where the four time-dependent functions are given by 
\begin{equation}
    h_{\nu}(t) = A(t)\begin{bmatrix}
        a(t) \cos\phase(t) \\ 
        b(t) \cos\phase(t) \\
        a(t) \sin\phase(t) \\
        b(t) \sin\phase(t)
    \end{bmatrix}
\end{equation}
and the corresponding time-independent amplitudes are
\begin{equation}
    c_{\nu} = \begin{bmatrix}
         \scri_{+} \cos\phi_0 \cos{2\psi} - \scri_{\times} \sin\phi_0 \sin{2\psi} \\
         \scri_{+} \cos\phi_0 \sin{2\psi} + \scri_{\times} \sin\phi_0 \cos{2\psi} \\
         -\scri_{+} \sin\phi_0 \cos{2\psi} - \scri_{\times} \cos\phi_0 \sin{2\psi} \\
         -\scri_{+} \sin\phi_0 \sin{2\psi}  + \scri_{\times} \cos\phi_0 \cos{2\psi}
    \end{bmatrix} \,.
\end{equation}
For a signal with constant amplitude \mbox{$A(t) = 1$},
these expressions reduce to the  parametrization commonly used in CW 
searches~\cite{Jaranowski:1998qm,Prix:2006wm,Wette:2011eu,Dreissigacker:2018afk}.

Designing a detection statistic à la Eq.~\eqref{eq:det_short} in this situation is
not so straightforward. The standard choice in the CW literature is the
$\mathcal{F}$-statistic~\cite{Jaranowski:1998qm,Cutler:2005hc}, 
which maximizes Eq.~\eqref{eq:logl} with respect to $c_{\nu}$,
\begin{equation}
    \mathcal{F} \propto \max_{c_0, c_1, c_2, c_3} \sum_{\nu=0}^{3} c_{\nu} \scal{x}{h_{\nu}}
    - \frac{1}{2} \sum_{\nu, \mu = 0}^{3} c_{\nu} c_{\mu} \scal{h_{\nu}}{h_{\mu}} \,.
\end{equation}
This choice provides a comparable performance to that of the optimal statistic 
for isotropically oriented signals~\cite{Searle:2008jv, Prix:2009tq}.
Other prescriptions have been proposed to deal with different signal and noise
hypotheses~\cite{Prix:2009tq,Dergachev:2011pd,Jaranowski:2010rn,Krishnan:2004sv,
Prix:2011qv,Whelan:2013xka,Bero:2018xyq,Covas:2022xyd,Prix:2024xpl,
Keitel:2015ova,Keitel:2013wga}. For the purposes of this work, we will directly compare
inner products in order to assess the accuracy of the SFT framework. The formulation
of a suitable detection statistic for long-duration early-warning applications is left to future work~\cite{Prix:2007zh,Keppel:2012ye,Prix:2024xpl,Deng:2025qhx}

Throughout this work, we neglect the Doppler modulation
of the detector on the signal for the sake of simplicity. As thoroughly shown in
Refs.~\cite{Jaranowski:1998qm,Prix:2005gx,Krishnan:2004sv, Patel:2009qe,
Dergachev:2011pd,2014PhRvD..90d2002A,fasttracks} (see~Ref.~\cite{Tenorio:2021wmz} 
for a review), the required machinery to account for this effect can be seamlessly
included into our framework and does not introduce major drawbacks
(see Sec.~\ref{sec:say_doppler}).

\section{Accelerated inner products\label{sec:sft_def}}

\subsection{Broad strategy}

With these ingredients, we now derive an efficient implementation of Eq.~\eqref{eq:fd_product}
for long-duration, inspiral-only GW signals. The main assumption is that, within a sufficiently
short time segment (to be formalized later), the Fourier transform of $h(t)$ is concentrated within
a narrow frequency range. This is similar to the assumptions required when using the
stationary phase approximation~\cite{Droz:1999qx}. 
We refer to such segmented signals as ``quasi-monochromatic.''

Note that any signal with a non-quasi-monochromatic component can be separated into two
segments so that the quasi-monochromatic one (which tends to last for longer) is analyzed
as in this paper while the rest (which tends to be significantly shorter) is analyzed
with standard methods. For a compact binary, which typically emits an 
inspiral-merger-ringdown signal, only the inspiral portion is taken as quasi-monochromatic.

Our goal is to formulate Eq.~\eqref{eq:fd_product} using SFTs
in a similar  manner to current implementations of the
\mbox{$\mathcal{F}$-statistic}~\cite{Williams:1999nt,Prix:2011qv,prixCFSv2}.
This approach is desirable for several reasons: 
\begin{enumerate} 
\item Fourier transforms of the data will be computed once
and will be valid for any template within a given parameter-space region. 
\item Noise
non-stationarities can be treated by whitening data on a per-SFT basis. 
\item Gaps in the data are automatically
accounted for, as they simply correspond to missing SFTs.
\item Waveform and detector quantities will be evaluated in the time domain, where
    they are more naturally defined, removing the need for approximations in the
        frequency domain.
\item The number of waveform evaluations
will be reduced significantly; rather than once per data sample, they will only be required once per SFT.
\end{enumerate}
Moreover, the use of SFTs  significantly reduces the memory footprint of a waveform evaluation,
allowing for their batch evaluation. This makes the SFT formulation suitable for GPU hardware. Computational implications are discussed in Sec.~\ref{sec:comp_impl}.

\subsection{Quasi-monochromatic signals\label{sec:say_doppler}}

We start by expressing Eq.~\eqref{eq:fd_product} in the time domain,
\begin{equation}
    \scal{d}{h} = \int_{0}^{T} \mathrm{d}t \, d(t) \, h(t) \,,
    \label{eq:td_product}
\end{equation}
where $T$ is the maximum time range so that the signal remains quasi-monochromatic.
The time series $d(t)$ entering Eq.~\eqref{eq:td_product} is the whitened observed data
\begin{equation}
    d(t) \doteq 4\,\mathrm{Re} \int_{0}^{\infty} \mathrm{d}f \frac{\tilde{x}^{*}(f)}{\Sn(f)} e^{-2 i \pi f t} \,.
    \label{eq:whitened_data}
\end{equation}
The process of whitening a strain dataset has been discussed at length
elsewhere~\cite{Finn:1992wt,
Jaranowski:1998qm, Krishnan:2004sv, Astone_2005, Allen:2005fk,Cannon:2011vi,
2014PhRvD..90d2002A, bsd, Usman:2015kfa, Tsukada:2017cuf, 2024EPJC...84.1023M, CabournDavies:2024hea}. 
We also assume that backgrounds containing
overlapping signals, such as those expected to affect next-generation detectors,
have been properly dealt with~\cite{Samajdar:2021egv,Pizzati:2021apa,Hourihane:2022doe,Alvey:2023naa}.

Data $d$ is measured on a discrete set of timestamps \mbox{$\{t_j = j \Delta t, j=0, \dots, N-1\}$}. 
We divide the observed data into $N_{\mathrm{SFT}}$ disjoint time segments
\mbox{$\mathcal{T}_{\alpha} = [t_{\alpha}, t_{\alpha+1}]$}.
Each of these segments has a duration of \mbox{$\Tsft = t_{\alpha + 1} - t_{\alpha}$} and contains
$n_{\mathrm{SFT}} = \Tsft / \Delta t$ samples.
The time $\Tsft$ should be chosen so that the GW  phase  within a segment can be Taylor-expanded to
second order (see Sec.~\ref{subsec:impl})
\begin{align}
    \phase(t) &\approx \phase_{\alpha}  + \Delta\phase_{\alpha}(t) \,, \\
    \Delta \phase_{\alpha}(t) &=   2 \pi f_{\alpha} (t - t_{\alpha}) + \pi \dot{f}_{\alpha} (t - t_{\alpha})^2 \,,
\end{align}
where the $\alpha$ subscript denotes evaluation at time $t_{\alpha}$ and 
\begin{align}
    f_{\alpha} = \frac{1}{2 \pi}
    \left. \frac{\mathrm{d}\phase}{\mathrm{d}t}  \right|_{t_{\alpha}} ,\,
    \dot{f}_{\alpha}  = \frac{1}{2 \pi}
    \left. \frac{\mathrm{d}^2\phase}{\mathrm{d}t^2}  \right|_{t_{\alpha}} \,.
\end{align}
This implies that for $t \in \mathcal{T}_{\alpha}$,
the waveform strain $h(t)$ is described by 4 numbers
\mbox{$\{A_{\alpha}, \phase_{\alpha}, f_{\alpha}, \dot{f}_{\alpha} \}$} rather than the initial
$n_{\mathrm{SFT}}$ samples. In particular one has
\begin{equation}
    h_{\alpha}(t) = \begin{cases}
    \amp_{\alpha} \proj_{\alpha} e^{
    i [\phase_{\alpha} + \Delta\phase_{\alpha}(t)]
    }
    & {\rm for} \quad t\in \mathcal{T}_{\alpha} \,,
    \\
   0 & {\rm for} \quad  t\notin \mathcal{T}_{\alpha}\,.
   \end{cases}
    \label{eq:h_tada}
\end{equation}
such that
\begin{equation}
    h(t) \approx \sum_{\alpha=0}^{N_{\mathrm{SFT}} - 1} h_{\alpha}(t)  \,.
\end{equation}
Let us further assume that the evolution of the signal amplitude
is slow enough to be approximated as a constant within a segment $\mathcal{T}_{\alpha}$.
This involves both the GW amplitude $A_{\alpha}$ and the detector amplitude modulation, 
here encapsulated in the projector $\Lambda_{\alpha}$. 
The case for $A_{\alpha}$ is trivial, as we are in the regime of validity of
the stationary phase approximation~\cite{Droz:1999qx}. The detector amplitude
modulation varies on a timescale of $\gtrsim 1 \,{\rm day}$ ($\gtrsim 1 \,{\rm year}$) for
ground-(space-)based detectors. 
These are usually longer than those allowed by the variation
of $A_{\alpha}$, so this is not an issue in practice; otherwise, $\Tsft$ needs be chosen
according to $\Lambda_{\alpha}$. 

Note that we neglected the Doppler effect by assuming that $\phase(t)$ is a function of
detector time $t$. The correct expression would include time delays depending on the
sky location of the source $\hat{n}$ to relate the arrival time at the detector $t$ 
to the emission time at the source $t_{\mathrm{s}}(t; \hat{n})$; 
$\phase(t_{\mathrm{s}}(t; \hat{n}))$. This function is known 
in closed form (see e.g., Ref.~\cite{Wette:2023dom}). The only difference would be that
the numerical values of phase-related quantities would now become 
\begin{equation}
    f_{\alpha} = \frac{1}{2 \pi} 
    \left. 
    \frac{\mathrm{d}\phase}{\mathrm{d}t_{\mathrm{s}}}
    \frac{\mathrm{d}t_{\mathrm{s}}}{\mathrm{d}t}
    \right|_{t_{\alpha}}
\end{equation}
and so on. In other words, the overall shape of Eq.~\eqref{eq:h_tada} is unaffected
and the only change is in the meaning of phase-related quantities. Note that the
magnitude of the Doppler enters as multiplicative factor. While in general this effect
limits $\Tsft$ down to a fraction of a few hours for CW signals at
$\gtrsim \SI{100}{\hertz}$~\cite{Krishnan:2004sv,prixCFSv2}, the low frequencies 
involved in this work make this a subdominant effect. 

Equation~\eqref{eq:td_product} can now be expressed as a sum over disjoint time segments,
\begin{align}
    \scal{d}{h} &= \sum_{\alpha=0}^{N_{\mathrm{SFT}}-1} 
    \amp_{\alpha} \proj_{\alpha}  e^{i \phase_{\alpha}} \scal{d_{\alpha}}{e^{i \Delta\phase_{\alpha}}} \,,
    \label{eq:prod_barbar} \\
    \scal{d_{\alpha}}{e^{i \Delta\phase_{\alpha}}}
    &= \int_{0}^{\Tsft} \mathrm{d}\tau \, 
    d_{\alpha}(\tau) e^{
    i (2 \pi f_{\alpha} \tau  + \pi \dot{f}_{\alpha} \tau^2)
    } \,,
    \label{eq:scal_coh}
\end{align}
where we took $d_{\alpha}(\tau) = d(t_{\alpha} + \tau)$. In general,
data will be evaluated at discrete times \mbox{$d_{\alpha}[j] = d(t_{\alpha} + \tau_{j})$ with $\tau_{j} = j \Delta t$}. This yields
\begin{equation}
    \scal{d_{\alpha}}{e^{i \Delta\phase_{\alpha}}} = 
    \Delta t \sum_{j=0}^{n_{\mathrm{SFT}}- 1} d_{\alpha}[j] e^{ i (2 \pi f_{\alpha} \tau_j  + \pi \dot{f}_{\alpha}  \tau_j^2) } \,.
    \label{eq:og_C}
\end{equation}

\subsection{SFTs and Fresnel kernel}

We seek an efficient implementation of Eq.~\eqref{eq:og_C}.
A first naïve attempt would be to interpret it as a discrete Fourier transform of a 
data segment
(i.e. an SFT) with respect to the frequency parameter $f_{\alpha}$. This approach,
would make the SFT dependent on the waveform parameters,
which would thus need to be recomputed for every waveform across either template banks for GW searches or
likelihood evaluations for GW parameter estimation. This would also require keeping the 
original dataset in memory, without any computing advantages. 

Instead, we re-express Eq.~\eqref{eq:og_C} in terms of the SFT of the data 
\begin{align}
    \cohsum(f_{\alpha}, \dot{f}_{\alpha}; \tilde{d}_{\alpha}) &\doteq \scal{d_{\alpha}}{e^{i \Delta\phase_{\alpha}}}\notag  \\
 &    =  \Delta f \sum_{k=k_{\mathrm{min}}}^{k_{\mathrm{max}}} \tilde{d}_{\alpha}^{*}[k] \, 
    \Fre(f_{\alpha}- f_k, \dot{f}_{\alpha})  \,,
    \label{eq:with_f_1}
\end{align}
where
\begin{align}
	\Delta f &= \Tsft^{-1} \,,\\
        \tilde{d}_{\alpha}[k] &= \Delta t \sum_{j=0}^{n_{\mathrm{SFT}}-1} d_{\alpha}[j] e^{-i 2 \pi \tau_j f_k} \,,
        \label{eq:d_SFT}\\
        \Fre(f_0, f_1) &= \Delta t \sum_{j=0}^{n_{\mathrm{SFT}}-1} 
            e^{i (2 \pi f_0\tau_j +  \pi f_1 \tau_{j}^2)} \,.
        \label{eq:fresnel_og}
\end{align}
Equation~\eqref{eq:d_SFT} is the SFT of a data taken within a time segment 
$\mathcal{T}_{\alpha}$. As long as $\Tsft$ is chosen appropriately,
these are independent of waveform parameters and can be computed only once. 

\begin{figure}
    \includegraphics[width=\columnwidth]{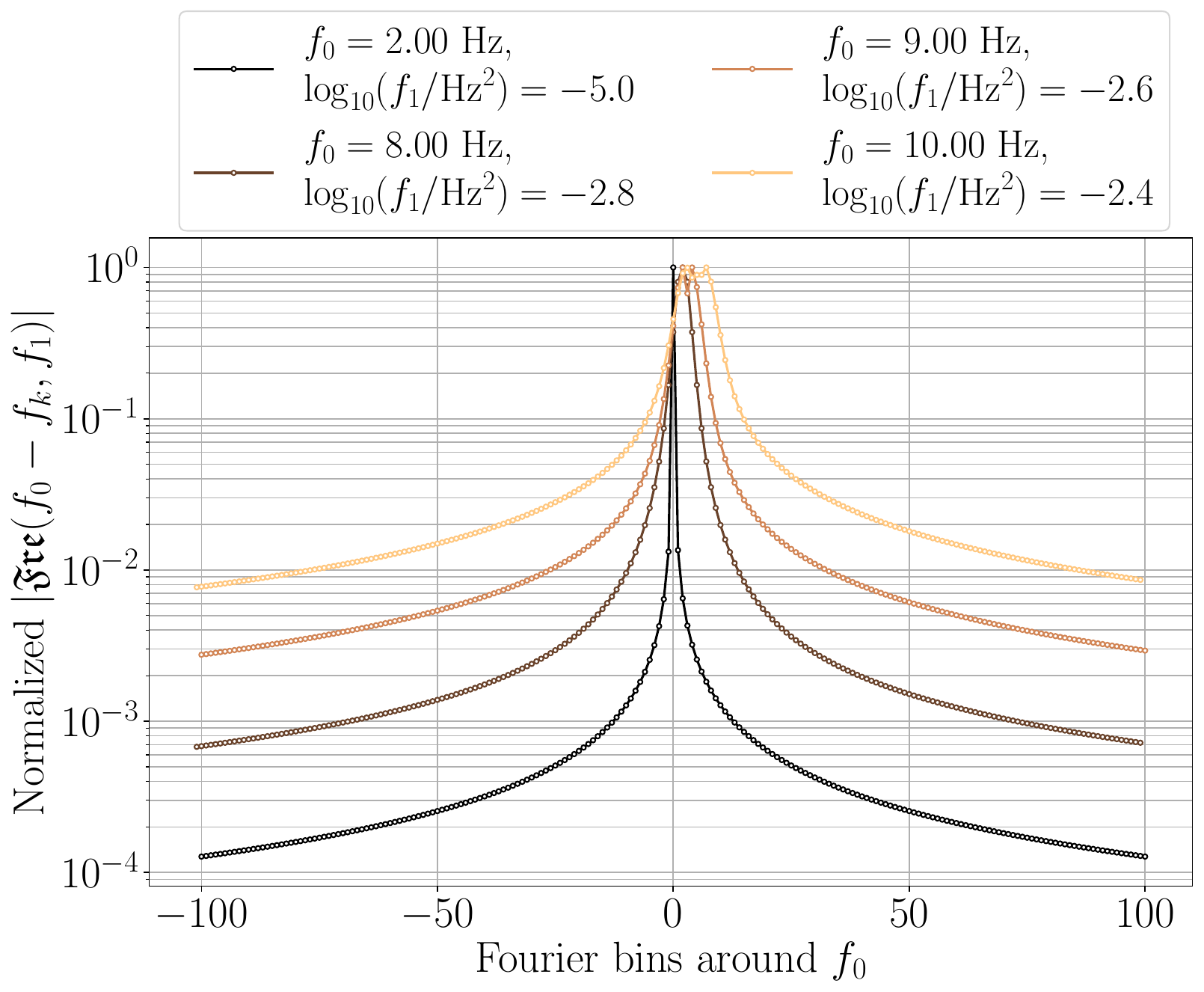}
    \caption{
        Magnitude of the kernel $\Fre$ normalized by its maximum value for different values of 
        $f_{0}, f_{1}$ taken from the evolution of a \mbox{$(1.4 - 1.4)\, \mathrm{M}_{\odot}$}
        BNS for $\Tsft = \SI{50}{\second}$. For small values of $f_1$, the signal is practically
        monochromatic and $\Fre$ narrows down to the Dirichlet kernel (see Appendix~\ref{app:dir}).
        For positive values of $f_1$, the signal visibly drifts towards higher 
        frequencies. 
    }
    \label{fig:Fre}
\end{figure}

The summation on the frequency index $k$ in Eq.~\eqref{eq:with_f_1} should run
over the full frequency spectrum. As shown in Fig.~\ref{fig:Fre}, however, 
the kernel $\Fre$ falls off rapidly for values of $f_0$ away from zero, 
allowing for the summation to be safely truncated to a finite range 
$k\in [k_{\mathrm{min}}, k_{\mathrm{max}}]$ around
\mbox{$f_{k} \sim f_{\alpha}$}.\footnote{Note that, if $f_{\alpha}$ is close to the
low end of the frequency band, the relevant range $[k_{\mathrm{min}}, k_{\mathrm{max}}]$
may include ``negative'' $k$ values corresponding to Fourier components indexed at 
$n_{\mathrm{SFT}} - k$ due to the inherent periodicity of the discrete Fourier transform.
}
This is comparable to the truncation of the Dirichlet kernel described
in~Refs.~\cite{Williams:1999nt,Allen:2002bp,prixCFSv2}.

Equation~\eqref{eq:fresnel_og} can be expressed in closed
form by taking the continuum limit:
\begin{equation}
    \Fre(f_0, f_1) =
    \int_{0}^{\Tsft} \, \mathrm{d}\tau \, e^{i (2 \pi f_0 \tau +  \pi f_1 \tau^2)} \,.
\end{equation}
This is justified as the kernel $\Fre$ does not involve any inherently discrete terms; in any case, the number of samples in 
our typical applications will be high enough to justify this step (see e.g.~Ref.~\cite{Allen:2002bp} as well as 
Appendix~\ref{app:dir} where a similar argument is made for a different kernel). 
Completing the square we find
\begin{equation}
    \Fre(f_{0}, f_1) = \frac{e^{-i \pi f_0^2 / f_1}}{\sqrt{2 f_1}} 
    \left\{
    C(u) - C(l) + i[ S(u) - S(l) ]
    \right\} \,,
\end{equation}
where
\begin{equation}
    \begin{aligned}
        l & = \sqrt{\frac{2}{f_1}} f_{0} \,,\\
        u & = l + \sqrt{2 f_1} \Tsft\,,
    \end{aligned} 
\end{equation}
and the Fresnel integrals are given by
\begin{equation}
    C(x) = \int_{0}^{x} \mathrm{d}\tau \cos{\left(\frac{\pi}{2} \tau^2\right)} \,,
\end{equation}
\begin{equation}
    S(x) = \int_{0}^{x} \mathrm{d}\tau \sin{\left(\frac{\pi}{2} \tau^2\right)} \,.
\end{equation}
For a monochromatic signal ($f_1 \rightarrow 0$) Eq.~\eqref{eq:fresnel_og}
reduces to the well-known Dirichlet kernel~\cite{Allen:2005fk} (see Appendix~\ref{app:dir}).

\subsection{SFT length\label{subsec:impl}}

The computations presented above require a suitable choice of  
$\Tsft$. To this end, we follow a well-established criterion in CW
searches~\cite{Krishnan:2004sv}, namely that the frequency evolution
described by Eq.~\eqref{eq:h_tada} deviates by less than a certain
fraction $\delta$ of the SFT frequency resolution $\Delta f$.
For a linear frequency evolution, the corresponding Taylor residual $R$ is bounded by
\begin{equation}
    \left| R \right| \leq \frac{1}{2} \Tsft^{2} \max_{\alpha} \ddot{f}_{\alpha}\,.
\end{equation}
The maximum allowed $\Tsft$ thus corresponds
to \mbox{$\left| R \right| = \delta \Delta f$}, which implies
\begin{equation}
    \Tsft(\delta)
    = \left(\frac{2 \delta}{\max_{\alpha}\ddot{f}_{\alpha}} \right)^{1/3} \,.
    \label{eq:max_tsft}
\end{equation}
For CBC signals, $\Tsft$ will be dominated by values of $\ddot{f}_{\alpha}$ at the end 
of the inspiral. This choice is not unique; rather, $\delta$ should be chosen such 
that the corresponding  $\Tsft$ is valid for as many waveforms as possible. If necessary,
different parameter-space regions can be analyzed using different values of $\Tsft$ in
order to maximize computational efficiency.

\subsection{Main takeaways}

The framework put forward in this paper corresponds to approximating
the inner product  $\scal{d}{h}$ with 
\begin{equation}
  \tdscal{d}{h} \doteq \Delta f \sum_{\alpha=0}^{N_{\mathrm{SFT}}-1}  
     A_{\alpha} 
    \proj_{\alpha} e^{i \phase_{\alpha}}
    \cohsum(f_{\alpha}, \dot{f}_{\alpha}; \tilde{d}_{\alpha}) \,.
    \label{eq:efficient}
\end{equation}
This formulation of the inner product can be intuitively interpreted as follows
(see Fig.~\ref{fig:example}).
First, the original dataset is processed into a set of SFTs, 
which we can think of as a 2D array with a time axis and a frequency axis
(more precisely, SFTs can be interpreted as a complex spectrogram,
which becomes a standard spectrogram after taking their squared absolute
value). To compute the inner product of a quasi-monochromatic signal, we follow
its instantaneous frequency as a function of time across the spectrogram,
and add the Fourier amplitudes coherently. From a computational point of view, this is a
coherent version of the CW search strategy put forward in
Ref.~\cite{fasttracks} by one of the authors.

Note that the data are the only quantity whose Fourier transform is ever computed.
All waveform quantities are evaluated in the time domain.

\section{Applications\label{sec:pilot}}

\subsection{Accuracy metric}

We now present two pilot applications of the SFT framework, considering early warnings for BNS in third-generation ground-based detectors and massive BBH in LISA. While these applications are mainly restricted to
CBC signals in their inspiral, we stress the methodology presented here can be combined
with other strategies to include further stages that break quasi-monochromaticity.

Concretely, we will assess the error due to using $\tdscal{d}{h}$ instead of $\scal{d}{h}$
using the relative error:
\begin{equation}
    r = 1 - \sqrt{\frac{\tdscal{d}{h}}{\scal{d}{h}}} \,.
\end{equation}
This quantity is comparable to the mismatch of a template bank~\cite{Owen:1995tm,Wette:2016raf,Allen:2019vcl, Allen:2021yuy},
which represents the fractional loss in detection statistic due to placing a finite number of
templates in a continuous space. The maximum allowed mismatch depends on the source under analysis,
with current CBC searches aiming for a maximum mismatch of
$1\%$ to $3\%$~\cite{Ajith:2007kx,Brown:2012nn,Sakon:2022ibh,Schmidt:2023gzj}.

The result of this analysis will be a quantification of the
loss produced by the approximations used in $\tdscal{\cdot}{\cdot}$
with respect to $\scal{\cdot}{\cdot}$. To do so, we generate a time series containing
a fiducial signal and compute $\scal{d}{h}$ by direct integration. We then compute $r$ for
different values of $\delta$ and \mbox{$\Delta k = k_{\mathrm{max}} - k_{\mathrm{min}}$}.
Acceptable setups will correspond to those so that $r \lesssim 1\%$.
In general, more permissive setups (higher $r$) yield computationally cheaper
implementations. 

\begin{table}
\renewcommand{\arraystretch}{1.3}
    \begin{tabular}{c| rrrrrr}
    $\delta$  & 3.0 & 2.0 & 1.0 \\
    \hline
        $\Tsft / \mathrm{s}$ & 106 & 92 & 73 \\
            $\Nsft$  & 4396 & 5066 & 6384 \\
    \end{tabular}
    \caption{
        SFT configurations for a BNS signal from \SI{1}{\hertz} to \SI{40}{\hertz}
        as described in Sec.~\ref{sec:bns}.
        The corresponding loss $r$ is shown in Fig.~\ref{fig:bns_mismatch}.
    }
    \label{table:toy_Tsft}
    \renewcommand{\arraystretch}{1}
\end{table}

\begin{figure}
    \includegraphics[width=\columnwidth]{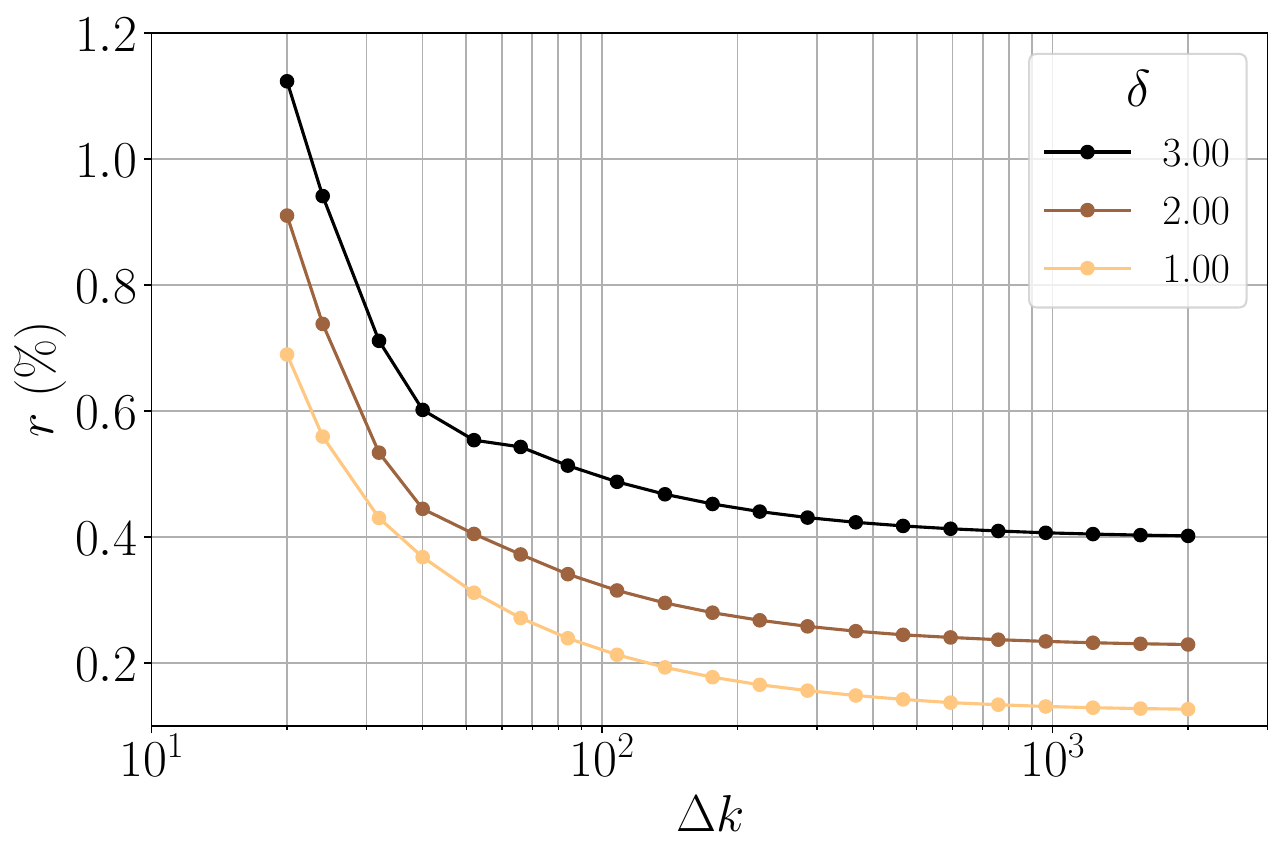}
    \caption{
        Relative error for a \mbox{$(1.4 - 1.4) \, \mathrm{M}_{\odot}$}
        BNS as described in Sec.~\ref{sec:bns} for different choices of $\delta$ and 
        $\Delta k$. 
    }
    \label{fig:bns_mismatch}
\end{figure}

\subsection{Early warning for BNS mergers in third-generation ground-based detectors\label{sec:bns}}

As a first application, we simulate an early-alert search for a BNS system.
The two objects have masses of $\SI{1.4}{\sun}$, corresponding to a chirp mass of
$\chirpM = \SI{1.21}{\sun}$. The signal is observed from $\SI{1}{\hertz}$ up
to \SI{10}{\hertz}, with a sampling frequency of \SI{40}{\hertz}. This frequency band
is consistent with future-generation detectors such as Einstein Telescope~\cite{ET:2019dnz}
and Cosmic Explorer~\cite{Reitze:2019iox}. This results in about \mbox{$N_0 \approx 2 \times 10^{7}$}
data points spanning a total of $130$ hours.

We assume the orientation and sky position of this source is compatible with the projector  
\begin{equation}
    \proj_{t} = [a(t) + b(t)](\mathrm{Re} + \mathrm{Im}) \,,
\end{equation}
where we took $c_{\nu} = 1$ and 
\begin{align}
    a(t) = \cos{\left(2 \pi t / 1\,\mathrm{day}\right)} \,,\\
    b(t) = \sin{\left(2 \pi t / 1\,\mathrm{day}\right)} \,,
\end{align}
which qualitatively reproduce the expected daily amplitude modulation due to the detector motion.
We numerically explore a range of $\delta$ as listed in Table~\ref{table:toy_Tsft}.
In this particular case,
\mbox{$\max_{\alpha} \ddot{f}_{\alpha} \approx \SI{5e-6}{\hertz/\second^2}$}.

As shown in Fig.~\ref{fig:bns_mismatch}, configurations with
\mbox{$\delta \lesssim 3$} and \mbox{$\Delta k \gtrsim 20$} 
yield acceptable relative errors (\mbox{$r < 1 \%$}). This is equivalent to a number of
SFTs of about \mbox{$\Nsft \sim 10^{3-4}$}. Compared to $N_0 \sim 10^{7}$, the number of points at which
the waveform must be evaluated has dropped by 3 to 4 orders of magnitude. In other terms, this corresponds
to diminishing the sampling frequency from $\SI{40}{\hertz}$ to $\Tsft^{-1} \sim (0.01 - 0.1) \mathrm{Hz}$.

\begin{table}
\renewcommand{\arraystretch}{1.3}
    \begin{tabular}{c|rrrrrr}
        $\delta$ & 3.0 & 2.0 & 1.0 \\
    \hline
        $\Tsft / \mathrm{s}$ & 2502 & 2185 & 1735 \\
        $\Nsft$ & 1540 & 1764 & 2222 \\
    \end{tabular}
    \caption{
        SFT configurations for a massive BBH signal from
        \SI{e-4}{\hertz} to \SI{e-3}{\hertz}
        as described in Sec.~\ref{sec:mbbh}.
        The corresponding loss $r$ is shown in Fig.~\ref{fig:mbbh_mismatch}.
    }
    \label{table:MBH_Tsft}
    \renewcommand{\arraystretch}{1}
\end{table}

\begin{figure}
    \includegraphics[width=\columnwidth]{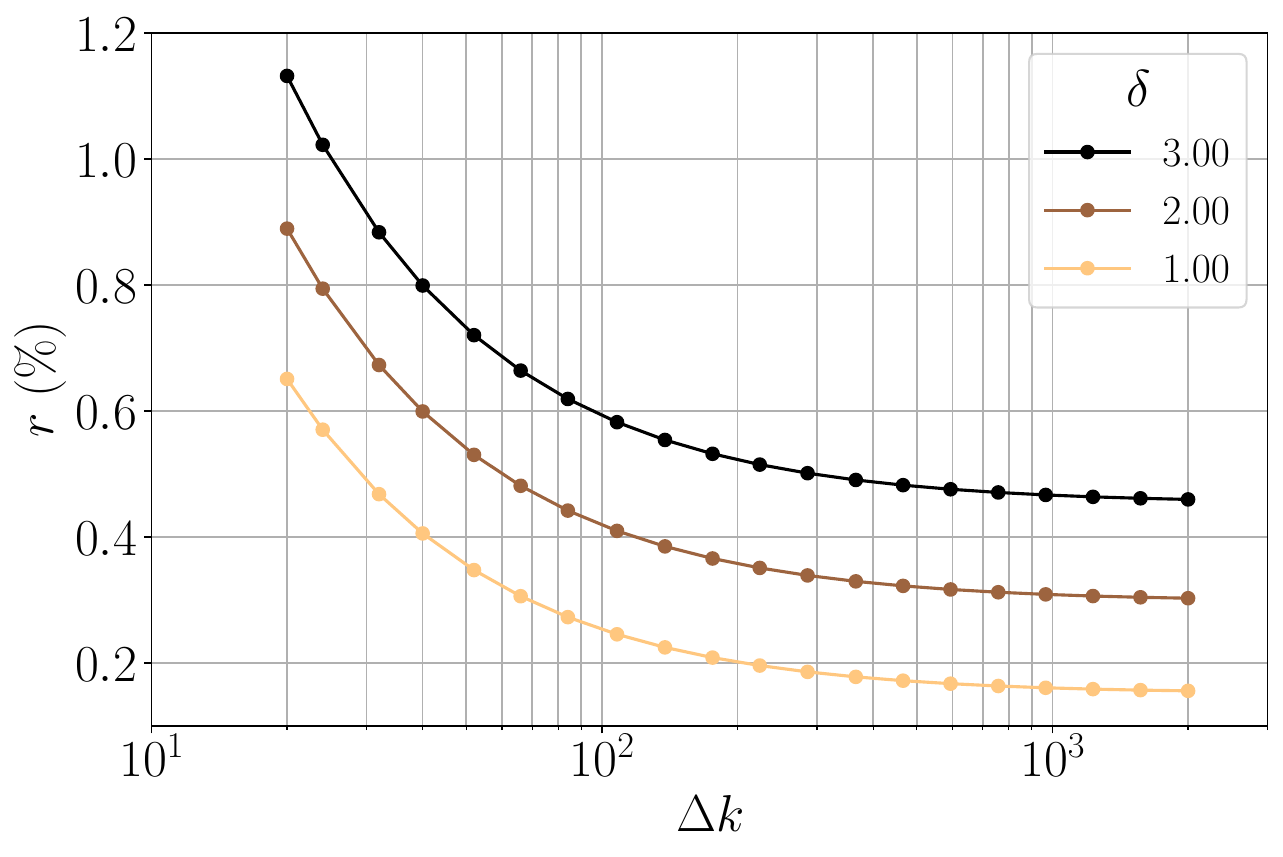}
    \caption{
        Relative error for a \mbox{$(10^6 - 10^6) \, \mathrm{M}_{\odot}$}
        massive BBH as described in Sec.~\ref{sec:mbbh} for different choices
        of $\delta$ and $\Delta k$.
    }
    \label{fig:mbbh_mismatch}
\end{figure}

\subsection{Early warning for massive BBH mergers in LISA\label{sec:mbbh}}

As a second application, we consider a massive BBH as observed by the space
detector LISA~\cite{LISA:2024hlh}. In this case, we take the two black holes 
to have a mass of $10^6 \mathrm{M}_\odot$ each and no spin. The system is observed 
from \SI{e-4}{\hertz}, corresponding to the start of the LISA sensitive band,
up to \SI{1.9e-3}{\hertz}, which is the frequency of the Minimum Energy Circular 
Orbit (MECO) of the system~\cite{Cabero:2016ayq}. This is the frequency at which
\mbox{\phenomt{}}~\cite{Estelles:2020osj,Estelles:2020twz,Estelles:2021gvs}
terminates the inspiral phase and, as a result, is a good proxy for the end of 
the quasi-monochromatic behavior. 
This signal lasts for about 45 days. At a sampling frequency of \SI{2}{\hertz}, 
which corresponds to the minimum acceptable sampling frequency covering the full
LISA sensitive band~\cite{LISA:2024hlh}, this corresponds to
 $N_0 \approx 7.5 \times 10^{6}$ samples.

The sub-millihertz frequency band is well in the regime of validity of the long-wavelength
approximation~\cite{Marsat:2020rtl}, which allows us to treat LISA as a LIGO-like
detector with a yearly amplitude modulation
\begin{align}
    a(t) = \cos{\left(2 \pi t / {1 \, \mathrm{year}}\right)} \,,\\
    b(t) = \sin{\left(2 \pi t / {1 \, \mathrm{year}}\right)} \,.
\end{align}
This is qualitatively correct, but note we are ignoring the Doppler modulation on
the waveform as previously mentioned.
Similarly to the previous example, we take $c_{\nu} = 1$.

As shown in Fig.~\ref{fig:mbbh_mismatch}, in this case $r \approx 1\%$ is achieved at
$\delta = 3$ and $\Delta k \sim 30$, which corresponds to
\mbox{$\Tsft \approx \SI{40}{\minute}$}
and \mbox{$N_{\mathrm{SFT}} \approx 1.5 \times 10^{3}$} (see Table~\ref{table:MBH_Tsft}). 
Once more, this is a reduction of about three orders of magnitude in the number of
time samples. 

Crucially, the overall data analysis strategy for LISA data will be qualitatively different from that of
LIGO/Virgo/KAGRA, since the sheer amount of sources in band calls for the use of a
global-fit strategy~\cite{Littenberg:2023xpl, Katz:2024oqg, Strub:2024kbe, Deng:2025wgk}.
Nevertheless, the framework we present here can still play a role in such analyses due to the
significant reduction in the required memory size and the advantages in dealing with non-stationarities
and gaps in the data compared to the use of bare time-domain or frequency-domain data.

\section{Computational implications\label{sec:comp_impl}}

There are three key computational advantages in
our implementation of the GW inner product. 

\subsection{SFTs\label{subsec:SFTs}}

The advantages of SFTs with respect to time-domain or frequency-domain representations have
been outlined in Sec.~\ref{sec:sft_def}. This data format is also advantageous from a computing
perspective.

In a GW experiment, data is naturally taken as a time series containing $N_0$ samples.
Obtaining a frequency domain representation of this dataset involves $\mathcal{O}(N_0 \log N_0)$
operations using a Fast Fourier Transform, and requires storing the complete dataset before
this operation can be performed.

Similarly, the cost of computing an SFT using the Fast Fourier Transform (FFT) scales as
$\mathcal{O}(n_{\mathrm{SFT}} \log n_{\mathrm{SFT}})$. These can be computed online and in parallel
as data arrives at the detector, allowing for analyses with very low latency.
Moreover, the cost of computing all the SFTs in a dataset scales as $\mathcal{O}(N_0 \log n_{\mathrm{SFT}})$, 
which is lower than computing the full transform by a factor $\log{N_0} / \log{n_{\mathrm{SFT}}} \sim 2$ according to the results in Sec.~\ref{sec:pilot}. 

This shows that SFTs are an appropriate and efficient representation, in a similar manner to
wavelets~\cite{Cornish:2020odn}, to analyze data streams with time-dependent
statistical properties.

\subsection{Scalar-product evaluation\label{subsec:single}}

Within the SFT framework, evaluating the approximate inner product $\tdscal{\cdot}{\cdot}$ involves two steps:
\begin{enumerate}
    \item Evaluate waveform quantities \mbox{$\{A_{\alpha}, \phase_{\alpha}, f_{\alpha}, \dot{f}_{\alpha}\}$} across SFTs.
    \item Weight and sum the relevant SFT bins according to Eq.~\eqref{eq:efficient}.
\end{enumerate}
Note that SFTs can be computed once in advance and then reused for a given parameter-space region.

Without the SFT framework, waveforms in step (i) must be evaluated at each of the $\mathcal{O}(N_0)$ time samples,
unless their implementation is capable of benefiting from an acceleration scheme~\cite{Vinciguerra:2017ngf,
Garcia-Quiros:2020qlt,Morisaki:2021ngj, Zackay:2018qdy,Leslie:2021ssu,Krishna:2023bug,Cornish:2021lje}.
The SFT framework, on the other hand, only requires to evaluate a time-domain waveform on
$\mathcal{O}(\Nsft)$ time samples. As shown in Sec.~\ref{sec:pilot}, $N_0 / \Nsft \sim 10^{3-4}$.
For waveforms such as 
\mbox{\phenomt{}}~\cite{Estelles:2021gvs,Estelles:2020twz,Estelles:2020osj},
which are implemented
using closed-form expressions, this implies the SFT framework outright reduces their computing cost and
memory requirements by three to four orders of magnitude. 

Step (ii) involves adding $\mathcal{O}(\Delta k \times \Nsft)$ bins after computing their corresponding weights
using $\Fre$ and $\proj_{\alpha}$. Results in Sec.~\ref{sec:pilot} show $\Delta k \sim 10^2$ to be an acceptable
value. As a result, in terms of number of operations, $\Delta k \times \Nsft \sim 10^{5-6}$ is one to two orders
of magnitude lower than the direct integration given by Eq.~\eqref{eq:fd_product}, which involves $N_0 \sim 10^7$
frequency bins.

Overall, given a closed-form time-domain waveform model, evaluating the SFT-based inner product
$\tdscal{\cdot}{\cdot}$ instead of $\scal{\cdot}{\cdot}$ results in a reduction of three to four
orders of magnitude in both waveform evaluation time and memory footprint, and a reduction of one
to two orders of magnitude in computing the inner product. 

In our implementation, we evaluate $\Fre$ using the Fresnel integral functions
available in~\jax{}~\cite{jax}. While efficient, this means $\Fre$ values are re-computed
from scratch whenever required, making their evaluation a dominant contribution of the
scalar product. As discussed in Sec.~\ref{subsec:gpu}, evaluation costs still end up
competitive with respect to other approaches. The development of an efficient evaluation
scheme of the $\Fre$ kernel is left for future work.

\subsection{Vectorized inner product\label{subsec:gpu}}

The gains in computing cost and memory consumption reported in the previous subsection suggest
GPU parallelization as a promising approach to further accelerate the evaluation of
Eq.~\eqref{eq:efficient}.

In this work, both \mbox{\phenomt{}} and $\tdscal{\cdot}{\cdot}$ have been 
implemented as part of the
\sfts{} package~\cite{sfts} using \jax{}~\cite{jax}, a high-level \textsc{python} front end
which allows for just-in-time compilation, GPU acceleration, and, crucially, automatic vectorization.

To parallelize Eq.~\eqref{eq:efficient} on a GPU, we use the \textsc{vmap} instruction in \jax{},
which transforms a waveform operating on a single set of parameters into a waveform operating
on batches of parameters. The implementation is otherwise unchanged.
The maximum batch size (i.e. how many waveforms are processed in parallel) of course depends on the computing capabilities of the machine at hand. Note that all waveforms generated
within a batch will have the same length; this is not a problem whenever using SFTs, as the
relevant waveform portion stops before merger.

To test the computing cost, we generate a batch of waveforms using our re-implementation 
of the inspiral portion of \mbox{\phenomt{}} by randomly sampling masses within 1\%
of the examples previously described and evaluating Eq.~\eqref{eq:efficient}. 
We perform all our tests using a
\mbox{13th Gen Intel(R) Core(TM) i7-1355U} CPU and a \mbox{NVIDIA H100 64GB} GPU.

\begin{enumerate}
\item
For the BNS case ($\delta = 1,\, \Delta k = 100$), the computing
cost of evaluating Eq.~\eqref{eq:efficient}
for a single waveform using a CPU is \SI{0.02}{\second}. 
Using \textsc{vmap} wcan evaluate Eq.~\eqref{eq:efficient} for a batch 
of 100 waveforms, with an average cost of $\SI{0.01}{\second}$ per waveform.
On a GPU, a single waveform can be evaluated in \SI{0.8}{\milli\second},
while for a batch of 1000 waveforms the average cost 
is \SI{0.4}{\milli\second} per waveform;

\item
For the massive BBH case ($\delta =3, \, \Delta k = 30$), Eq.~\eqref{eq:efficient} 
for a single waveform on a CPU takes \SI{0.01}{\second}. The batch size on a CPU in 
this case can be increased up to 1000 waveforms, yielding an average cost of
\SI{3}{\milli\second} per waveform. On a GPU, a single waveform can be evaluated in
\SI{0.8}{\milli\second}, while for a batch of 1000 waveforms the average computing
cost drops to \SI{0.08}{\milli\second} per waveform.
\end{enumerate}
These GPU applications, which have been made possible by the SFT framework,
demonstrate a further computing cost reduction of two to three orders of magnitude due
to \jax{}'s \textsc{vmap} primitive with virtually no development cost. Altogether,
whenever applicable, the SFT framework results in a reduction of three to five orders
of magnitude in computing Eq.~\eqref{eq:fd_product} with respect to
non-optimized formulations.  These results are complementary to those
of Ref.~\cite{GQ2025}, which evaluate a single waveform including the merger and
ringdown phases on a GPU.

\section{Conclusion}

We presented a new framework based on  SFTs to analyze
long-duration GW signals emitted by compact objects in binary systems. This is
one of the key data-analysis challenges posed by next-generation GW
detectors~\cite{ET:2019dnz,Reitze:2019iox,LISA:2024hlh}. 

The basic idea, shown in Fig.~\ref{fig:example}, is to Fourier transform short disjoint
segments (producing SFTs) so that waveform templates can be filtered against the relevant
portion of the data spectrum at different times in their evolution. This approach is inspired by CW search methods~\cite{Williams:1999nt,Prix:2011qv,prixCFSv2} and provides
several computational advantages at a negligible development cost.

Since SFTs are disjoint in time, noise non-stationarities can be dealt with on a per-SFT
basis~\cite{Finn:1992wt,Jaranowski:1998qm,Krishnan:2004sv, Astone_2005, Allen:2005fk,Cannon:2011vi,
2014PhRvD..90d2002A,Usman:2015kfa, CabournDavies:2024hea}. In a similar fashion, gaps in the data
simply correspond to missing SFTs. Finally, as thoroughly shown in the CW
literature~\cite{Tenorio:2021wmz,Riles:2022wwz,Wette:2023dom}, amplitude and frequency modulations 
caused by the GW detector can be accounted for in time domain on a per-SFT basis as well, 
vastly reducing the complexity of the analyses with respect to frequency-domain approaches \cite{Chen:2024kdc}. SFTs can be computed on-line as data arrives and is more efficient than
generating the full-time Fourier transform. Furhtermore, 
they can be recycled  for a broad region of the parameter space,
further amortizing their minimal implementation cost. In this aspect,
they offer comparable advantages to other time-frequency methods~\cite{Cornish:2020odn}.

In addition, SFTs allow for a reduction of three to five orders of magnitude in the computing cost of the
inner product [Eq.~\eqref{eq:fd_product}] for inspiral waveforms, which is the most fundamental
quantity in any GW data analysis routine~\cite{Finn:1992wt,1987thyg.book..330T}, with
respect to a non-optimized approach. This gain can be
explained by two components:
\begin{enumerate}
\item
First, the phase evolution of an inspiral waveform within an SFT can be
approximated in closed form by a quadratic Taylor expansion. This reduces the effective sampling
frequency of a waveform by three to four orders of magnitude and the computing cost of the inner
product by one to two orders of magnitude. 
\item Second, the reduced sampling frequency, combined with the
inherently parallel evaluation of Eq.~\eqref{eq:fd_product} using the SFT framework
[Eq.~\eqref{eq:td_product}], allow for the parallelization of multiple waveforms using a GPU.
This lowers the computing cost of Eq.~\eqref{eq:td_product} further by two to three orders of magnitude.
\end{enumerate}
Overall, we find a cost on the order of \mbox{$0.1 \, \mathrm{ms}$}
per inner product evaluation for next-generation GW analysis, 
depending on the specific application, using current GPUs.
As discussed in e.g. Ref.~\cite{Wang:2023tle, Miller:2023rnn,Hu:2024mvn}, template banks,
and in general the computing cost of GW data-analysis routines,
may grow by several orders of magnitude for next-generation detectors compared to LIGO/Virgo/KAGRA.
This makes the methods presented in this work critical to exploit the scientific
capabilities of those GW observatories. 

Although the computing advantages we achieved focus on inspiral-dominated signals,
the SFT framework can be seamlessly combined with standard methods to tackle the merger and
ringdown portions of the waveform. Due to their relatively shorter duration, these are less affected
by the difficulties encountered in the inspiral (see e.g. Ref~\cite{Marsat:2020rtl}).

While we have limited this presentation to use a single detector and a single GW mode, the
SFT framework can operate with multiple detectors and GW modes by extending the
definition of the projector operator, much like in other CW
analysis pipelines.

The SFT framework crucially relies on evaluating waveform amplitudes, phases, frequencies,
and frequency derivatives at arbitrary times. This requirement is incompatible with the current
interface exposed by the LIGO/Virgo/KAGRA Algorithm Library~\cite{lalsuite,swiglal}; as a result, we re-implemented
the closed-form time-domain inspiral portion of~\phenomt{} ~\cite{Estelles:2020osj,Estelles:2020twz,Estelles:2021gvs} 
using \jax{} to allow for automatic parallelization and differentiation operations,
which we release under the  \sfts{}~\cite{sfts} package together with the required tools to operate
within the SFT framework. This release complements other efforts in the community focused on frequency-domain
models~\cite{Edwards:2023sak,ml4gw} and cements vectorized waveforms as a crucial tool for the acceleration
of GW data-analysis workflows~\cite{fasttracks}.

The SFT framework, however, is not limited to closed-form waveform approximants. Any waveform
family~\cite{Varma:2019csw,Ramos-Buades:2023ehm,Nagar:2024dzj} capable of providing the required ingredients
can benefit from the computing advantages here discussed. Future work will investigate extensions and limitations
of the SFT framework to operate on precessing and/or eccentric waveform models, as well
as other more intricate sources such as extreme mass-ratio inspirals~\cite{Babak:2017tow}. These kinds of phenomena
are often described in the time-domain~\cite{GQ2025}, which is precisely the input domain expected
by the SFT framework.

In addition, the SFT framework can serve as a basis to further accelerate alternative strategies,
such as the semicoherent methods proposed for BNS early-warning searches~\cite{Miller:2023rnn},
stellar-mass BBHs in LISA~\cite{Bandopadhyay:2024lwv}, extreme mass-ratio
inspirals~\cite{Wen:2005xn,Gair:2006nj,Gair:2007bz, Ye:2023lok}, or long transient GW from young neutron
stars~\cite{Grace:2023kqq}.
Moreover, using Fresnel integrals may reduce the number of required SFTs in a CW search, further
lowering the cost of their semicoherent approaches~\cite{prixCFSv2,Prix:2011qv}.

Taken together, these results provide a promising solution to the problem of analyzing the
long-duration inspirals observed by next-generation detectors including the effect of
noise non-stationaries, data gaps, and modulations induced by the detector motion.

\section*{Acknowledgements}

We thank Ssohrab Borhanian, Arianna I. Renzini, Héctor Estellés, Jorge Valencia, and Sascha Husa for discussions.
R.T. and D.G. are supported by
ERC Starting Grant No.~945155--GWmining, 
Cariplo Foundation Grant No.~2021-0555, 
MUR PRIN Grant No.~2022-Z9X4XS, 
MUR Grant ``Progetto Dipartimenti di Eccellenza 2023-2027'' (BiCoQ),
and the ICSC National Research Centre funded by NextGenerationEU.
D.G. is supported by MSCA Fellowship No.~101064542--StochRewind and No.~101149270--ProtoBH.
Computational work was performed at CINECA with allocations 
through INFN and Bicocca, at MareNostrum5 with technical support
provided by the Barcelona Supercomputing Center (RES-FI-2024-3-0013),
and Artemisa with funding by the European Regional Development Fund and
the Comunitat Valenciana and technical support provided by the Instituto de Fisica Corpuscular (CSIC-UV).

\appendix

\section{Vectorized inspiral time-domain waveform\label{app:phenomt}}

The LIGO/Virgo/KAGRA Algorithm Library (\lalsuite{})~\cite{lalsuite, swiglal} provides a 
\textsc{Python} interface to several waveform models available
in the literature. Due to its generality, however, it is difficult to freely access 
the inner workings of a waveform model. For example, in the case of
\mbox{\phenomt{}}~\cite{Estelles:2020osj, Estelles:2020twz, Estelles:2021gvs}
it is not possible to evaluate $h_{+, \times}$ at an arbitrary time array
even though $h_{+, \times}$ are defined as closed-form functions of time.
This not only prevents sample-efficient algorithms (such as the one presented here),
but also needlessly complicates the use of GPU-parallelization through a
\textsc{python} front-end such as \jax{}~\cite{jax} or \textsc{pytorch}~\cite{pytorch}.

For this reason, we have re-implemented the inspiral part of the dominant $(2, \pm 2)$ 
mode of the \mbox{\phenomt{}} approximant using \textsc{jax}. 
Our implementation is distributed via the \sfts{} \cite{sfts} package. 

Two main advantages result from our re-implementation. 
First, \jax{}'s automatic differentiation
simplifies the computation of $\dot{f}_{\alpha}$ and $\ddot{f}_{\alpha}$ using the
closed-form implementation of $\phase(t)$ or $f(t)$. Second, \jax{}'s \textsc{vmap}
primitive allows for the vectorization of $h_{+, \times}$ with respect to both
time and waveform parameters. When combined with just-in-time compilation and
GPU-support, this allows for an unprecedented speed up in waveform evaluation as discussed
in Sec.~\ref{sec:comp_impl}.

\section{Linear phase drift \label{app:dir}}

The case of a monochromatic signal (i.e. $\dot{f}_\alpha = 0$) was thoroughly discussed
in Ref.~\cite{Allen:2002bp}. The procedure is similar to the $\dot{f}_{\alpha} \neq 0$ case:
\begin{equation}
    \cohsum(f_{\alpha}; \tilde{d}_{\alpha}) 
    = \Delta f \sum_{k=k_{\mathrm{min}}}^{k_{\mathrm{max}}}
    \tilde{d}^{*}[k] \, \Dir(f_{\alpha}- f_{k}) \,,
\end{equation}
where $\Dir$ is the Dirichlet Kernel
\begin{equation}
    \Dir(f) = \Delta t \sum_{j=0}^{n_{\mathrm{SFT}}- 1} e^{i 2 \pi \tau_j f} 
    = n_{\mathrm{SFT}} e^{i \pi \Tsft f} \mathrm{sinc}(\Tsft f) \,,
\end{equation}
where
\mbox{$\mathrm{sinc}(x) = \sin{(\pi x)}/(\pi x)$} and the latest equality holds in the 
limit \mbox{$n_{\mathrm{SFT}} \gg 1$}. Comparing with Eq.~\eqref{eq:fresnel_og}, $\Fre$ 
reduces to $\Dir$ in the limit $f_1 = 0$.

The $\Dir$  kernel drops off
significantly already at \mbox{$\Delta k  \lesssim 5$} \cite{Allen:2002bp}.
This is consistent with the early stages
of a BNS in Fig.~\ref{fig:bns_mismatch}; see also Fig.~\ref{fig:Fre},
where shorter $\Tsft$ values produced narrower $\Fre$ kernels as the signals within
the SFTs became closer to monochromatic ones.

\bibliography{references}

\end{document}